\documentclass[preprint,superscriptaddress,showpacs,preprintnumbers,amsmath,amssymb]{revtex4}

\usepackage{graphicx}
\usepackage{dcolumn}
\usepackage{bm}
\usepackage{color}

\begin{document}

\title{Stability of Strutinsky shell correction energy\\ in relativistic mean field theory
\footnote{Supported by the National Key Basic Research Programme of
China under Grant No 2007CB815000, the National Natural Science
Foundation of China under Grant Nos 10435010, 10775004 and
10221003.}}

\author{Yifei Niu}
 \affiliation{State Key Lab Nucl. Phys. {\rm\&} Tech., School of Physics, Peking University, Beijing 100871, China}

\author{Haozhao Liang}
 \affiliation{State Key Lab Nucl. Phys. {\rm\&} Tech., School of Physics, Peking University, Beijing 100871, China}
 \affiliation{Institut de Physique Nucl\'eaire, IN2P3-CNRS and Universit\'e Paris-Sud,
    F-91406 Orsay Cedex, France}

\author{Jie Meng \footnote{Email: mengj@pku.edu.cn }}
 \affiliation{State Key Lab Nucl. Phys. {\rm\&} Tech., School of Physics, Peking University, Beijing 100871, China}
 \affiliation{Department of Physics, University of Stellenbosch, Stellenbosch, South Africa}

\date{\today}

\begin{abstract}
The single-particle spectrum obtained from the relativistic mean
field (RMF) theory is used to extract the shell correction energy
with the Strutinsky method. Considering the delicate balance between
the plateau condition in the Strutinsky smoothing procedure and the
convergence for the total binding energy, the proper space sizes
used in solving the RMF equations are investigated in detail by
taking $^{208}$Pb as an example. With the proper space sizes, almost
the same shell correction energies are obtained by solving the RMF
equations either on basis space or in coordinate space.

\end{abstract}


\maketitle

The liquid drop model of the nucleus was historically the first
model to be proposed as an explanation of the properties of the
nucleus. Then the occurrence of the magic numbers has been one of
the strongest evidences for the nuclear shells. The shell effect
presents a fluctuation in the binding energy, the so-called shell
correction energy, which can be supplemented to the liquid drop
model to improve the descriptions of the nuclear masses. One of the
most practical and effective methods for extracting the shell
correction energy was proposed by Strutinsky
\cite{Strutinsky67,Strutinsky68}. The Strutinsky method has been
widely applied to the calculations of the masses, shapes and fission
barriers of nuclei \cite{Moller95, Bolsterli72}.

Two key points in the Strutinsky smoothing procedure are the
so-called plateau condition as well as the reliable single-particle
spectrum.

On one hand, the plateau condition means that the shell correction
energy should be insensitive to the smoothing range $\gamma$ and the
order of generalized Laguerre polynomial $M$, since these two
variables are only parameters in the smoothing procedure which have
no physical meanings \cite{Strutinsky68,Brack73}. The plateau
condition has been investigated in the past several decades. The
Strutinsky method was first combined with the Nilsson model. In this
case a good plateau could appear due to the infinity of the
potential \cite{Strutinsky67}. However, no plateau appears in the
realistic finite potential because of the discontinuity of the level
density at the upper edge of the potential \cite{Brack73}. A
possible solution for this problem is to complete the spectrum
outside the bound states region, using the resonances in the
continuum region \cite{Ross72} or using the positive eigenvalues
obtained by diagonalizing the Hamiltonian in a harmonic-oscillator
basis \cite{Bolsterli72, Nazarewicz94}. The first method is
cumbersome, especially for deformed nuclei, although the plateau
could appear if resonances up to 60 MeV are included \cite{Ross72}.
The second one is more practical, however, the shell correction
energy and even the appearance of the plateau condition depend on
the number of harmonic oscillator shells included in the
basis\cite{Bolsterli72, Nazarewicz94}. This requires a delicate
adjustment to the shell number $N_0$. In the non-relativistic
framework, the plateau condition for $^{208}$Pb has been checked
with the phenomenological folded Yukawa potential\cite{Bolsterli72,
Moller95}, where the shell correction energy was found to be stable
in the range of $8\leq N_{0}\leq 13$, thus $N_{0}=12$ was
recommended and a good plateau was found.

On the other hand, the self-consistent relativistic mean field (RMF)
theory\cite{Serot86}, which has received wide attention due to its
successful description of lots of nuclear
phenomena\cite{Ring96,Vretenar05,Meng06}, could provide the
single-particle spectrum microscopically. With this single-particle
spectrum thus obtained, the shell correction energies in superheavy
nucleus have been investigated and it is found that the shell
correction energy at the saddle point is too important to be
neglected\cite{ZhangWei03}, which will influence the synthesis of
the superheavy nuclei\cite{LiWF06,LiuZH06}.

The RMF equations can be solved by diagonalizing the Hamiltonian in
a harmonic-oscillator basis space\cite{Gambhir90} or by solving the
coupled channel differential equations within a finite box $[0,R]$
in coordinate space\cite{Horowitz81}, meanwhile the single-particle
spectrum outside the bound states region can also be obtained. In
this way, the level density in the continuum depends on the space
size, i.e. the shell number for fermions $N_f$ (bosons $N_b$) or the
box size $R$. Therefore, the space size used in solving the RMF
equations is one of the crucial quantities for the Strutinsky
smoothing procedure. Meanwhile, the choice of the space size should
guarantee the convergence for the total binding energy.

In this Letter, based on the single-particle spectrum obtained from
the RMF theory, the shell correction energy will be extracted by the
Strutinsky method. The efforts will be focused on determining the
proper space size used in solving the RMF equations via the delicate
balance between the plateau condition and the convergence for the
total binding energy.

The basic ansatz of the RMF theory is a Lagrangian density where
nucleons are described as Dirac spinors that interact via the
exchange of several mesons (the scalar $\sigma$, the vector
$\omega,$ and isovector vector $\rho$) and the photon\cite{Serot86}.

The classical variation principle gives equations of motion for the
nucleon, mesons and the photon. As in many applications, we study
the ground-state properties of nuclei with time reversal symmetry,
thus the nucleon spinors are the eigenvectors of the stationary
Dirac equation
 \begin{equation}
 \label{dirac}
 \left[\boldsymbol{\alpha}\cdot
 \boldsymbol{p}+V(\boldsymbol{r})+\beta(M+S(\boldsymbol{r}))\right]\psi_i
            =\epsilon_i\psi_i,
 \end{equation}
 and equations of motion for mesons and photon are
 \begin{eqnarray}
 \label{KG}
 \begin{array}{l} \left(-\triangle + m_\sigma^2\right)\sigma(\boldsymbol{r})
  =-g_\sigma\rho_s(\boldsymbol{r})-g_2\sigma^2-g_3\sigma^3,\\
 (-\triangle+m_\omega^2)\omega^0(\boldsymbol{r})
                       =g_\omega \rho_v(\boldsymbol{r}),\\
 (-\triangle+m_\rho^2)\rho^0(\boldsymbol{r})=
                                 g_\rho\rho_3(\boldsymbol{r}),\\
  -\triangle A^0(\boldsymbol{r})=e\rho_p(\boldsymbol{r}),
  \end{array}
  \end{eqnarray}
  where $\omega^0$
and $A^0$ are timelike components of the vector $\omega$ and the
photon fields and $\rho^0$ the third component of the timelike
component of the isovector vector $\rho$ meson. Eq.~(\ref{dirac})
and Eq.~(\ref{KG}) are coupled to each other by the vector and
scalar potentials
 \begin{eqnarray}
 \begin{array}{l}
 \displaystyle V(\boldsymbol{r})= g_\omega\omega^0
            +g_\rho\tau_3\rho^0
            +e\frac{1-\tau_3}{2}A^0,\\
 \displaystyle S(\boldsymbol{r})= g_\sigma \sigma,
 \end{array}
 \end{eqnarray}
and various densities
 \begin{eqnarray}
 \begin{array}{l}
 \displaystyle\rho_s(\boldsymbol{r})=\sum_{i=1}^A \bar{\psi}_i\psi_i,\\
 \displaystyle\rho_v(\boldsymbol{r})=\sum_{i=1}^A \psi_i^\dagger\psi_i,\\
 \displaystyle\rho_3(\boldsymbol{r})=\sum_{i=1}^A \psi_i^\dagger\tau_3\psi_i,\\
 \displaystyle\rho_c(\boldsymbol{r})=\sum_{i=1}^A
 \displaystyle\psi_i^\dagger\frac{1-\tau_3}{2}\psi_i.
 \end{array}
 \end{eqnarray}

The above RMF equations can be solved in basis space or coordinate
space as mentioned above. The single-particle spectrum thus obtained
will be used to calculate the shell correction energies with the
Strutinsky method.

  The Strutinsky method is based on the assumption that the
  realistic shell correction energy can be extracted uniquely from
  a mean field model.
  The shell correction energy, representing the fluctuating part of
  the binding energy, is defined as the difference between the total
  single-particle energy and its smooth part \cite{Brack73},
  \begin{equation}
   E_{\rm shell}=E-\widetilde{E}=\sum_{i=1}^{N(Z)}\epsilon_i
   -2\int_{-\infty}^{\widetilde{\lambda}}\epsilon\widetilde{g}(\epsilon)d\epsilon,
  \end{equation}
  where $N$ ($Z$) is the neutron (proton) number, $\epsilon_i$ is the single-particle
  energy.
  The smoothed Fermi level $\tilde{\lambda}$ is determined by the conservation of the
  particle number,
  $\displaystyle N(Z)=2\int_{-\infty}^{\tilde{\lambda}}\tilde{g}(\epsilon)d\epsilon$. The
  smoothed level density $\tilde{g}(\epsilon)$ takes the form,
  \begin{eqnarray}
   \tilde{g}(\epsilon)&=&\frac{1}{\gamma}\int_{-\infty}^{\infty}
   \left(\sum_{i=1}^{\infty}\delta(\epsilon'-\epsilon)\right)f(\frac{\epsilon'-\epsilon}{\gamma})d\epsilon'\nonumber\\
   &=&\frac{1}{\gamma}\sum_{i=1}^{\infty}f(\frac{\epsilon_i-\epsilon}{\gamma}).
  \end{eqnarray}
with the smoothing range $\gamma$. The folding function is usually
taken as $\displaystyle f(x)=\frac{1}{\sqrt{\pi}}e^{-x^2}P(x)$,
where $P(x)$ is an generalized Laguerre polynomial
$L_{M}^{1/2}(x^2)$. In practice, a cutoff is needed in the summation
of the single-particle energy, and the shell correction energy is
stable when the cutoff energy is large enough.

The Strutinsky method, as shown above, has two additional variables,
the smoothing range $\gamma$ and the order of generalized Laguerre
polynomial $M$. Since neither $\gamma$ nor $M$ has physical meaning,
the value of the shell correction should be insensitive to these
quantities within a certain range of values, i.e. the so-called
plateau condition,
    \begin{equation}
     \frac{\partial E_{\rm shell}}{\partial \gamma}=0,\qquad  \frac{\partial E_{\rm shell}}{\partial
     M}=0.
    \end{equation}
This forms an important criterion for the feasibility of the
Strutinsky method.

Taking the nucleus $^{208}$Pb as an example, the neutron
single-particle spectrum is calculated by the RMF theory with the
parameter set PK1\cite{LongWH04}. In the Strutinsky smoothing
procedure, the unit of the smoothing range $\gamma$ is adopted as
$\hbar\omega_0=41A^{-1/3}(1+\frac{1}{3}\frac{N-Z}{A})$~MeV, and the
energy cutoff is 30~MeV.

First of all, solving the RMF equations in basis space, it is found
that the shell correction energies are almost independent on the
shell number for bosons $N_b$, which could be understood as follows.
The shell correction energies are determined by the single-particle
level density. Since the single-particle energies are the
eigenvalues of the Hamiltonian matrix for fermions (nucleons),
changing the shell number for fermions $N_f$ will change the
single-particle level density in the positive energy region, whereas
changing $N_b$ doesn't influence it. Thus, the shell correction
energies are not sensitive to $N_b$. In the following investigation,
$N_b=20$ is fixed as a proper number \cite{Ring97}, and the efforts
will be focused on the influence of $N_f$ on the shell correction
energies.

\begin{figure*}
\includegraphics[width=5 in]{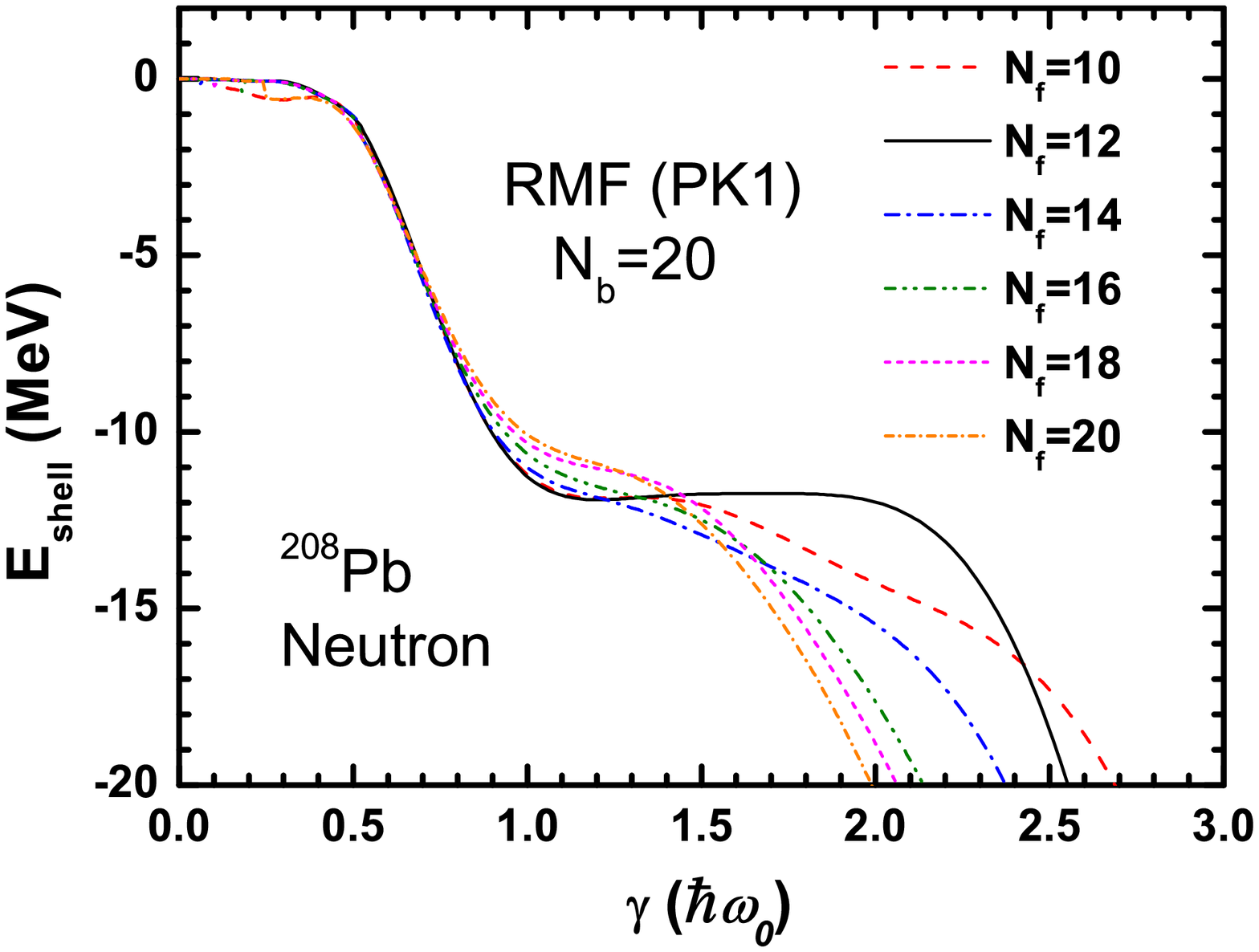}
\includegraphics[width=5 in]{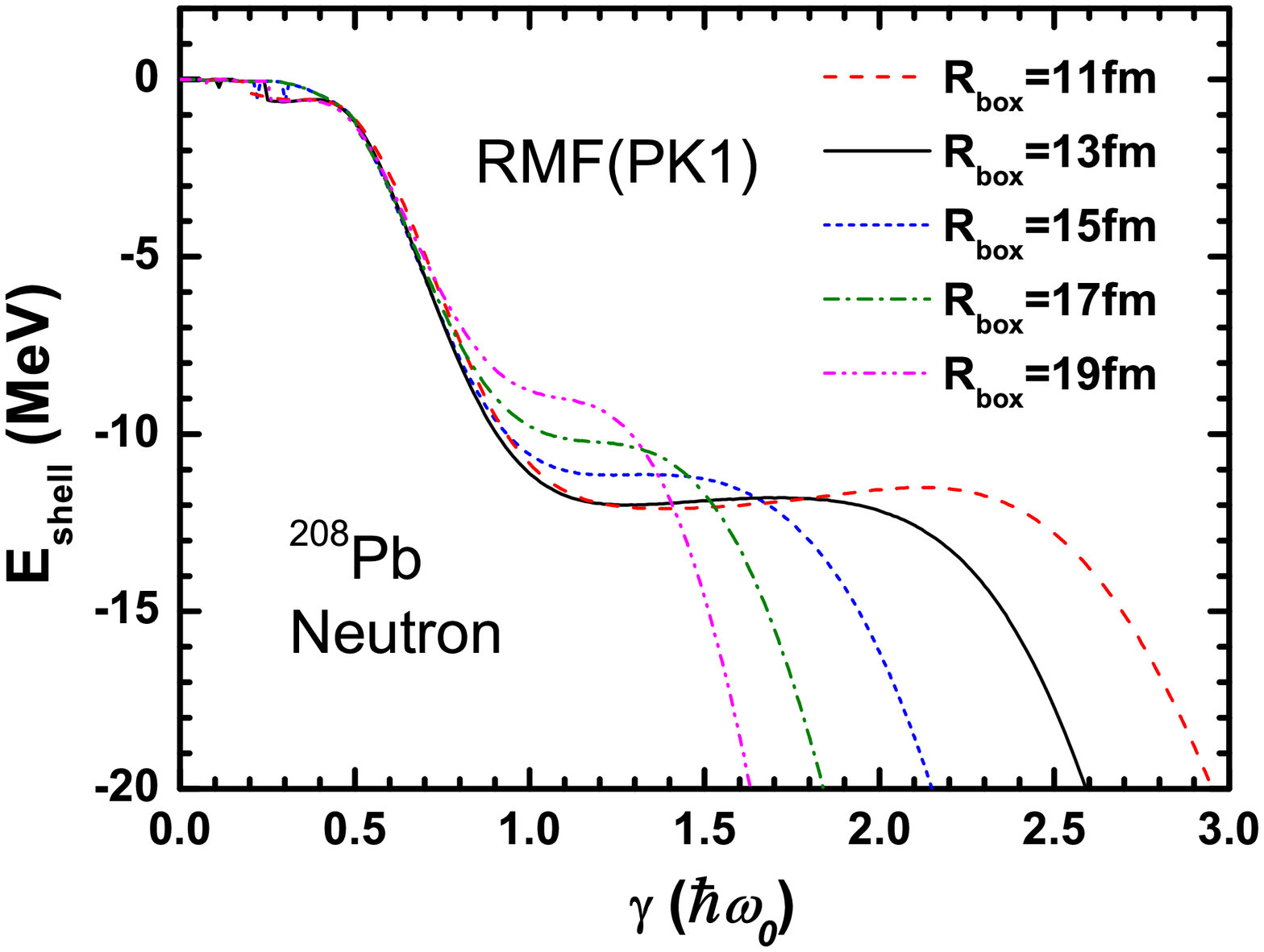}
\caption{ \label{fig:gamma}Neutron shell correction energies as a
function of the smoothing range $\gamma$ for $^{208}$Pb calculated
by RMF theory with PK1 parameter.
              The RMF equations are solved in basis space with the shell numbers $N_{f}=10,12,14,16,18,20$ for fermions
              and the fixed shell number $N_b=20$ for bosons (the upper panel),
              as well as in coordinate space with the box sizes $R_{\rm box}=11, 13, 15, 17, 19$ fm (the lower panel), respectively.
}
\end{figure*}

In the upper panel of Fig.~\ref{fig:gamma}, the neutron shell
correction energies as a function of the smoothing range $\gamma$
for the nucleus $^{208}$Pb are shown, where the RMF equations are
solved in basis space with different shell numbers $N_{f}=10 \sim
20$ for fermions, and with fixed shell number $N_b=20$ for bosons. A
well-pronounced plateau is seen in the case of $N_{f}=12$. It is
also found that the plateau vanishes when $N_f>12$. The reason is
similar as the case not including the positive energy part
\cite{Ross72}. When $N_f$ increases, the single-particle levels in
the positive energy region become much denser than those in the
bound region. The discontinuity in the single-particle level density
makes the plateau vanish.

Another important criterion for choosing the shell numbers is that
$N_f$ should be large enough to achieve the convergence for the
total binding energy. According to this point, $N_f\geq 12$ was
proposed \cite{Ring97}. Specifically, in the present calculation,
the accuracy for the total binding energy is 0.04$\%$ when $N_f=12$
is chosen.

Taking the balance between the above criteria, the size of basis
space used in solving the RMF equations should be taken as $N_f=12$
and $N_b=20$, which is in accordance with the space size for
folded-Yukawa potential \cite{Bolsterli72}. The neutron shell
correction energy reads $-11.92 \sim -11.74$~MeV within the range
$1.2\leq \gamma\leq 1.8$.

In the lower panel of Fig.~\ref{fig:gamma}, the neutron shell
correction energies as a function of the smoothing range $\gamma$
for the nucleus $^{208}$Pb are shown, where the RMF equations are
solved in coordinate space with different box sizes $R_{\rm box}=11
\sim 19$ fm. The plateau condition is well fulfilled in the cases of
$R_{\rm box}\leq 15$ fm. Especially, the shell correction energies
are stable between $R_{\rm box}=11$ fm and $R_{\rm box}=13$ fm. It
is also found that the plateau vanishes gradually as $R_{\rm box}>
15$ fm. The reason is the same as that in the basis space.

On the other hand, $R_{\rm box}$ should be large enough to achieve
the convergence for the total binding energy. It is found that the
accuracy for the total binding energy is 0.01$\%$ when $R_{\rm
box}=13$ fm is chosen.

Therefore, based on the above two criteria, the box size $R_{\rm
box}=13$~fm is recommended for solving the RMF equations in
coordinate space. The neutron shell correction energy reads $-12.00
\sim -11.80$~MeV within the range $1.2\leq \gamma\leq 1.8$.

In order to verify the choice of the above space sizes, the
dependence of the plateau condition on the order of generalized
Laguerre polynomial $M$ will be investigated.

\begin{figure*}
 \includegraphics[width=5 in]{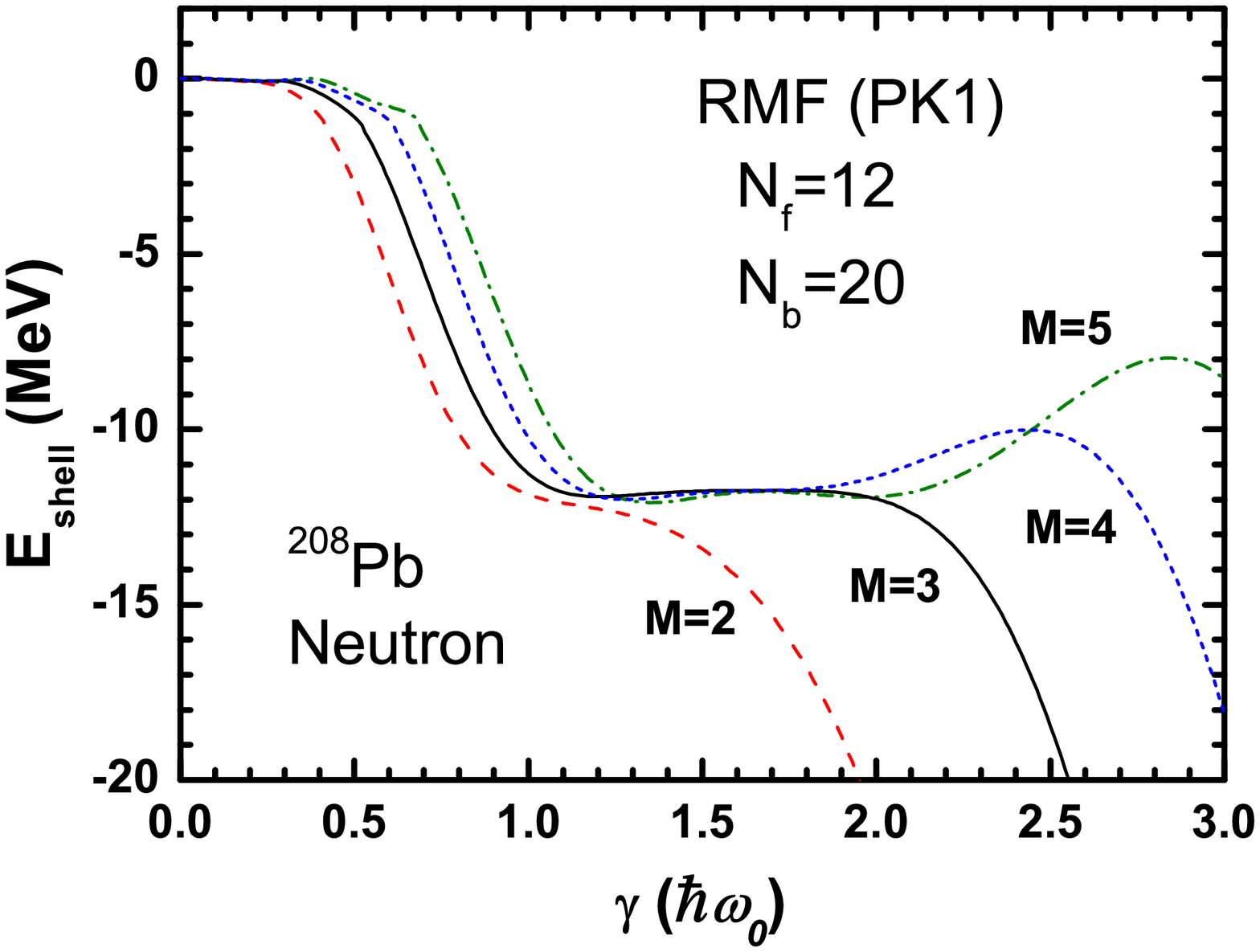}
 \includegraphics[width=5 in]{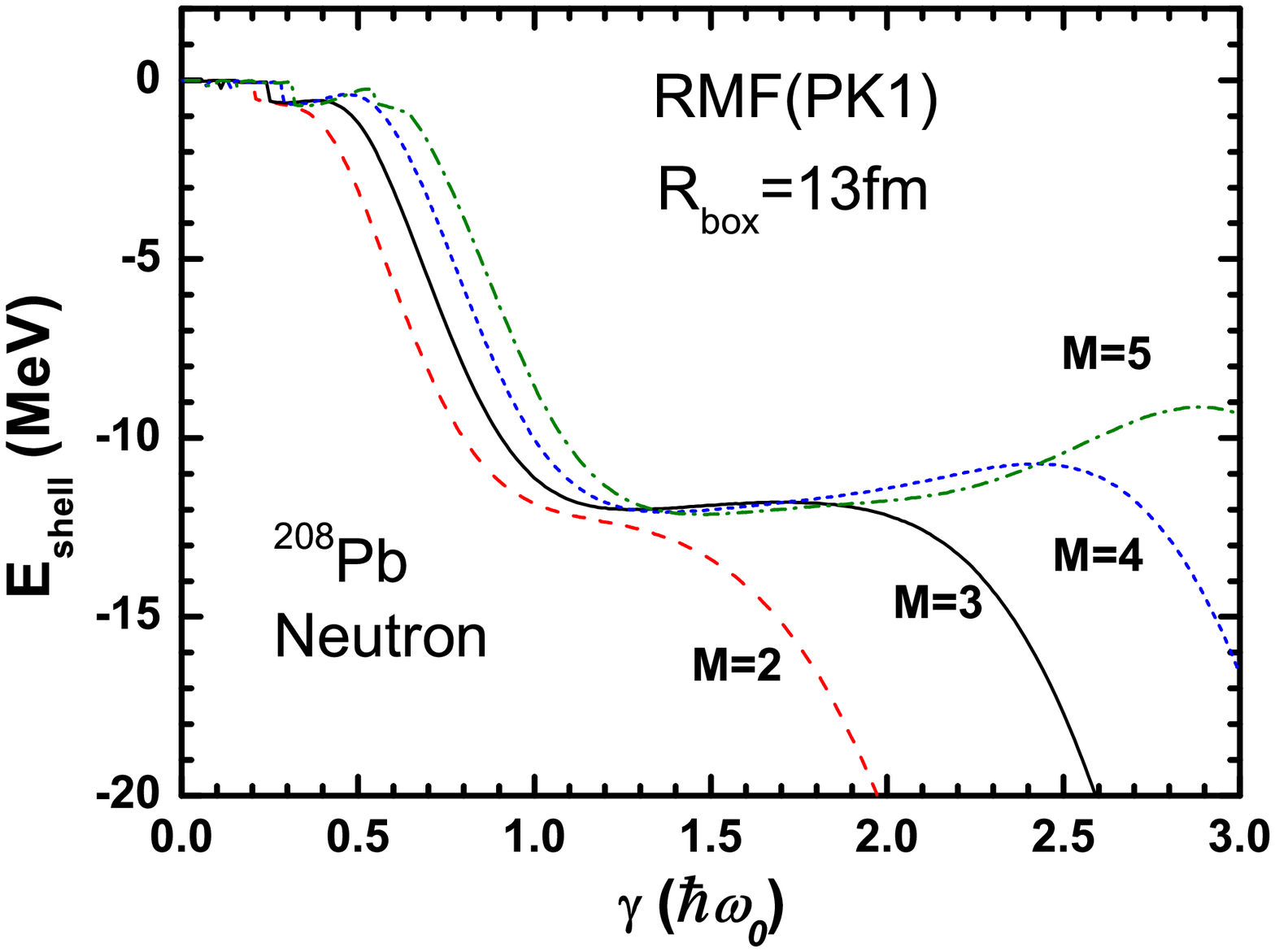}
 \caption{\label{fig:Mgamma} Neutron shell correction energies as a
function of the smoothing range $\gamma$
              for $^{208}$Pb calculated by RMF theory with PK1 parameter.
              The RMF equations are solved in basis space with the shell numbers $N_{f}=12, N_b=20$ (the upper panel),
              as well as in coordinate space with the box size $R_{\rm box}=13$ fm (the lower panel).
              The four different curves correspond to the order $M=2,3,4,5$ of generalized Laguerre polynomial, respectively.}
\end{figure*}

In Fig.~\ref{fig:Mgamma}, the neutron shell correction energies as a
function of the smoothing range $\gamma$ for the nucleus $^{208}$Pb
are shown. The RMF equations are solved in basis space with the
shell numbers $N_{f}=12, N_b=20$ (the upper panel), as well as in
coordinate space with the box size $R_{\rm box}=13$ fm (the lower
panel). The four different curves correspond to the order
$M=2,3,4,5$ of generalized Laguerre polynomial, respectively. It is
found that there is no clear plateau in the case of $M=2$, whereas
well-pronounced plateaus appear in the cases of $M=3,4,5$, and they
are almost identical. Thus, the shell correction energies calculated
with the above selected space sizes are independent on the order $M$
of generalized Laguerre polynomial within a reasonable range.
Furthermore, $M=3$ is an optimal value for the present calculations,
since the optimal value of $M$ is reached as soon as the shell
correction energy remains constant when $M$ is increased
\cite{Brack73}.

In summary, considering the delicate balance between the plateau
condition in the Strutinsky smoothing procedure and the convergence
for the total binding energy, the proper space sizes used in solving
the RMF equations are investigated in detail by taking $^{208}$Pb as
an example. Solving the RMF equations in basis space, the shell
number for fermions $N_f=12$ is recommended, while it is found that
the shell correction energies are almost independent on the shell
number for bosons $N_b$. Meanwhile, $R_{\rm box}=13$~fm is
recommended when the RMF equations are solved in coordinate space.
The above choices of the space sizes lead to almost identical
neutron shell correction energies $E_{\rm shell}\sim -11.90$~MeV in
both cases, and are verified by the independence of the shell
correction energy on the order $M$ of generalized Laguerre
polynomial around $M=3$.

\newpage

\begin{thebibliography}{99}

\bibitem{Strutinsky67} Strutinsky V M 1967 {\em Nucl. Phys.} A~{\bf 95} 420

\bibitem{Strutinsky68} Strutinsky V M 1968 {\em Nucl. Phys.} A~{\bf 122} 1

\bibitem{Moller95}Moller P et al 1995 {\em At. Data. Nucl. Data Tables}~{\bf 59} 185

\bibitem{Bolsterli72}Bolsterli M et al 1972 {\em Phys. Rev.} C~{\bf 5} 1050

\bibitem{Brack73} Brack M and Pauli H C 1973 {\em Nucl. Phys.} A~{\bf 207} 401

\bibitem{Ross72}Ross C K and Bhaduri R K 1972 {\em Nucl. Phys.} A~{\bf 188} 566

\bibitem{Nazarewicz94}Nazarewicz W, Werner T R and Dobaczewski J 1994 {\em Phys. Rev.} C~{\bf 50} 2860

\bibitem{Serot86} Serot B D and Walecka J D 1986 {\em Adv. Nucl. Phys.} {\bf 16} 1

\bibitem{Ring96} Ring P 1996 {\em Prog. Part. Nucl. Phys.}~{\bf 37} 193

\bibitem{Vretenar05} Vretenar D et al 2005 {\em Phys. Rep.} {\bf 409} 101

\bibitem{Meng06} Meng J et al 2006 {\em Prog. Part. Nucl. Phys.}~{\bf 57} 470

\bibitem{ZhangWei03} Zhang W et al 2003 {\em Chin. Phys. Lett.}~{\bf 20} 1694

\bibitem{LiuZH06} Liu Z H and Bao J D 2006 {\em Phys. Rev.} C~{\bf 74} 057602

\bibitem{LiWF06} Li W F et al 2006 {\em J. Phys. G: Nucl. Part. Phys.} {\bf 32} 1143

\bibitem{Gambhir90} Gambhir Y K, Ring P and Thimet A 1990 {\em Ann. Phys. (N.Y.)}~{\bf 198} 132

\bibitem{Horowitz81} Horowitz C J and Serot B D 1981 {\em Nucl. Phys.} A~{\bf 368} 503

\bibitem{LongWH04}Long W H et al 2004 {\em Phys. Rev.} C~{\bf 69} 034319

\bibitem{Ring97}Ring P, Gambhir Y K and Lalazissis G A 1997 {\em Comput. Phys. Commun.} {\bf 105} 77

\end{thebibliography}
\end{document}